\documentclass{article}

\usepackage{arxiv}

\usepackage[utf8]{inputenc} 
\usepackage[T1]{fontenc}    
\usepackage{url}            
\usepackage{booktabs}       
\usepackage{amsfonts}       
\usepackage{nicefrac}       
\usepackage{microtype}      
\usepackage{lipsum}		
\usepackage{graphicx}
\usepackage{doi}
\usepackage{amsmath}
\usepackage{amssymb}
\usepackage{graphicx}
\usepackage{caption}
\usepackage{xcolor}
\usepackage{makecell}
\usepackage{multirow}
\usepackage{multicol}
\usepackage{enumitem}
\usepackage{tikz}
\usepackage{hyperref}
\usepackage{biblatex}
\usepackage{pgfplots} 
\pgfplotsset{compat=1.16}

\usepackage{mathtools} 

\hypersetup{
    colorlinks,
    linkcolor={blue!80!black},
    citecolor={blue!80!black},
    urlcolor={blue!80!black}
}
\usepackage[caption=false]{subfig}
\definecolor{azure}{rgb}{0.0, 0.5, 1.0}
\definecolor{darkgreen}{rgb}{0.0, 0.5, 0.0}
\definecolor{amaranth}{rgb}{0.9, 0.17, 0.31}
\definecolor{cadetgrey}{rgb}{0.57, 0.64, 0.69}
\definecolor{aureolin}{rgb}{0.99, 0.93, 0.0}
\usepackage{etoolbox}

\newcommand{\tikzxmark}{%
\tikz[scale=0.23] {
    \draw[line width=0.7,line cap=round] (0,0) to [bend left=6] (1,1);
    \draw[line width=0.7,line cap=round] (0.2,0.95) to [bend right=3] (0.8,0.05);
}}
\newcommand{\tikzcmark}{%
\tikz[scale=0.23] {
    \draw[line width=0.7,line cap=round] (0.25,0) to [bend left=10] (1,1);
    \draw[line width=0.8,line cap=round] (0,0.35) to [bend right=1] (0.23,0);
}}

\usepackage{listings}

\newcommand{\bash}[1]{\lstinline[language=bash,breaklines=true]{#1}}
\newcommand{\cpp}[1]{\lstinline[language=c++,breaklines=true]{#1}}

\lstdefinelanguage{rust}
{
    keywords={
    true,false,
    unsafe,async,await,move,
    use,pub,crate,super,self,mod,
    struct,enum,fn,const,static,let,mut,ref,type,impl,dyn,trait,where,as,
    break,continue,if,else,while,for,loop,match,return,yield,in
    },
    ndkeywords={
    bool,u8,u16,u32,u64,u128,i8,i16,i32,i64,i128,char,str,Self,Option,Some,None,
    Result,Ok,Err,String,Box,Vec,Rc,Arc,Mutex,Cell,RefCell,HashMap,BTreeMap,macro_rules
    }
}
\newcommand{\rust}[1]{\lstinline[language=Rust,,breaklines=true]{#1}}

\title{Evaluating AI-generated code for C\texttt{++}, Fortran, Go, Java, Julia, Matlab, Python, R, and Rust}

\date{} 					

\author{	
	\href{https://orcid.org/0000-0003-3922-8419}{\includegraphics[scale=0.06]{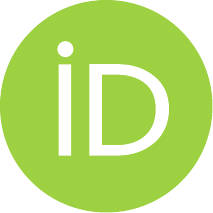}\hspace{1mm}Patrick Diehl} \\
	Louisiana State University\\
	Baton Rouge, 70803, LA, USA \\
        Applied Computer Science, \\
        Los Alamos National Laboratory\\
	\texttt{pdiehl@cct.lsu.edu} \\
         \And
       \href{https://orcid.org/0009-0000-4687-1416}{\includegraphics[scale=0.06]{orcid.pdf}\hspace{1mm}Noujoud Nader} \\
	Louisiana State University\\
	Baton Rouge, 70803, LA, USA \\
	\texttt{nnader@lsu.edu} \\
        \And        
        \href{https://orcid.org/0000-0002-7979-2906}{\includegraphics[scale=0.06]{orcid.pdf}\hspace{1mm}Steve Brandt} \\
	Louisiana State University\\
	Baton Rouge, 70803, LA, USA \\
	\texttt{sbrandt@@cct.lsu.edu} \\
        \And
        \href{https://orcid.org/0000-0002-8712-2806}{\includegraphics[scale=0.06]{orcid.pdf}\hspace{1mm}Hartmut Kaiser} \\
	Louisiana State University\\
	Baton Rouge, 70803, LA, USA \\
	\texttt{hkaiser@@cct.lsu.edu} \\
 }

\renewcommand{\shorttitle}{Evaluating AI-generated code for C\texttt{++}, Fortran, Go, Java, Julia, Matlab, Python, R, and Rust}

\begin{document}
\maketitle
\begin{abstract}
This study evaluates the capabilities of ChatGPT versions 3.5 and 4 in generating code across a diverse range of programming languages. Our objective is to assess the effectiveness of these AI models for generating scientific programs. To this end, we asked ChatGPT to generate three distinct codes: a simple numerical integration, a conjugate gradient solver, and a parallel 1D stencil-based heat equation solver. The focus of our analysis was on the compilation, runtime performance, and accuracy of the codes. 
While both versions of ChatGPT successfully created codes that compiled and ran (with some help), some languages were easier for the AI to use than others (possibly because of the size of the training sets used). Parallel codes---even the simple example we chose to study here---are also difficult for the AI to generate correctly.
\end{abstract}


\section{Introduction}
Generative AI algorithms have been used to create codes and high-quality content quickly and cost-effectively, tailored to specific user needs or criteria~\cite{wu2023ai,wang2023survey}.
This new tool has the potential to revolutionize traditional programming methodologies and change the way code is developed~\cite{brown2020language}. GitHub\ Copilot \cite{dakhel2023github}, for instance, is enhancing the developer experience by introducing features like Copilot Chat~\cite{openaiOpenAICodex}.
AI code generation tools typically operate on a tiered subscription model, with monthly per-user prices ranging from \$$10$ to \$$25$ at the time of this writing. Many providers also offer a free version tailored for individual users. The lower-priced tiers generally provide only basic functionality such as manual code generation through prompts and autocompletion. More advanced features, including model tuning and security scanning, are typically reserved for the more expensive tiers~\cite{trustradiusListCode}.
    
\noindent As these tools become more integrated into everyday coding practices, they pose both opportunities and challenges that merit careful examination and ongoing research. Recent advancements in AI-generated code represent a continuation of the challenges facing traditional computer science education. The emergence of AI code generators, coupled with the abundant availability of online code and platforms that enable contract cheating, starkly contrasts with the traditional image of programmers diligently crafting bespoke code~\cite{becker2023programming}. In addition, while these tools offer the potential to simplify the coding process and make programming more accessible to beginners, it is essential to recognize their limitations. These tools may not consistently generate code that is correct, efficient, or easy to maintain. Consequently, it is critical to thoroughly evaluate multiple factors when considering the adoption of such tools for coding tasks~\cite{zhang2023practices}. 

\noindent In November 2022, OpenAI introduced ChatGPT, a derivative of the generative pre-trained transformer (GPT), a large language model (LLM) based on transformer architecture that can comprehend human languages and generate text resembling human writing. Within just five days after its launch, the platform had already registered over one million users~\cite{taecharungroj2023can}. OpenAI improved the program on March 14th, 2023, with the release of GPT $4$, which promised better reasoning ability. In this study, we systematically evaluate the accuracy and efficacy of the code generated by ChatGPT both versions $3.5$ and $4$ in many programming languages. Our analysis focuses on the compilation and correctness of the generated code. Through this study, we aim to establish a comprehensive understanding of the AI-code generator ChatGPT's capabilities and its limitations. We selected programming languages from among the Top\ 20 in the TIOBE Index~\cite{tiobeTIOBEIndex}. The languages chosen for our analysis include C\texttt{++}, Fortran, Go, Java, Julia, Matlab, Python, R, and Rust.
The first example in this study is an example of numerical integration; the second, a conjugate gradient solver that incorporates matrix and vector operations; and lastly, a parallel 1D stencil-based heat equation solver~\cite{diehl2023benchmarking}. This last model problem was chosen due to its relevance in testing parallel computational workflows and its common usage in scientific computing. 

The paper is structured as follows: In Section~\ref{sec:related_work}, we will provide an overview of related work. In Section~\ref{sec:methodology}, we will present our methodology. The quality of the generated code and code metrics will be described in Sections~\ref{sec:quality} and~\ref{sec:code_metrics}. In Section~\ref{sec:conclusion}, we conclude our work and indicate possible future steps.

\section{Related Work}
\label{sec:related_work}
Previous studies evaluated the performance of AI models for code generation using different strategies. Chen at al.\ \cite{chen2021evaluating} released HumanEval, a new evaluation set measuring functional correctness for synthesizing programs from docstrings. They find that HumanEval solves $28.8$\% of the problems while GPT-$3$ solves $0$\%. Nguyen et al.\ in \cite{nguyen2022empirical} conducted an empirical investigation to assess the correctness and understandabilty of Copilot's proposed code. 
They found that Copilot's suggestions for Java had the highest correctness rate at $57$\%, whereas JavaScript had the lowest correctness rate at $27$\%. Another study evaluated the performance of an AI tool named Codex in localizing and fixing bugs~\cite{pearce2021can,prenner2022can}.
Instead of generating code, Tang et al. used eye tracking and IDE tracking~\cite{tang2023empirical} to conduct a user study to assess how programmers deal with errors when using Copilot.
Recently, Liu et al.\ \cite{liu2023refining} systematically studied the quality of $4,066$ ChatGPT-generated code implemented in Java and Python, for $2,033$ programming tasks.
They first analyzed the correctness of ChatGPT when generating code and found that while it could generate functional code, there were a number of quality errors. Providing the AI with feedback and allowing it to self-repair was of limited effectiveness in removing these quality issues.

These studies, we think, show the potential and promise for AI code generation, but also highlight its present weaknesses.
\section{Methodology}
\label{sec:methodology}
We use three numerical examples: numerical integration (\textbf{NI}), a conjugate gradient solver (\textbf{CGS}), and a parallel one-dimensional heat equation solver (\textbf{PHS}) using finite differencing. The code complexity increases with each example. We used the free version of ChatGPT\ $3.5$ and the paid version ChatGPT\ $4.0$ for our study. We used ChatGPT to generate the codes on \textit{02/27/2024} for the last example and \textit{06/05/2024} for the others.
The following queries were used to obtain the source code;
\begin{enumerate}
    \item Write a \textbf{language} code to compute the area between $-\pi$ and $2/3\pi$ for $\sin(x)$ and validate it.\\
    Here, we want to validate if ChatGPT can write a code to evaluate 
    \begin{equation}
        \int\limits_{-\pi}^{2/3\pi}  \sin(x) dx \text{.}
    \end{equation}
    \item Write a conjugate gradient solver in \textbf{language} to solve $A$ times $x$ equals $b$ and validate it.\\
    Here, we want to evaluate whether ChatGPT can write a conjugate gradient solver and apply it to a linear system of equations, i.e.
    \begin{align}
        A^{n\times n} \cdot x^n = b^n \text{ with } n\in\mathbb{Z}^+, A=A^T, \text{and } x^T A x > 0, \text{for all } X \in \mathbb{R}^n \text{.}
    \end{align}
    For more details about the conjugate gradient solver, we refer to~\cite{shewchuk1994introduction}.
    \item Write a parallel 1D heat equation solver using finite differencing in \textbf{language}.\\
    Here, we want to evaluate whether ChatGPT can write the code to solve
    \begin{align}
       \frac{\partial u}{\partial t} = \alpha\frac{\partial^2 u}{\partial x^2}, \quad 0 \leq x < L, t>0
    \end{align}
    where $\alpha$ is the material's diffusivity. For the discretization in space a finite difference scheme 
    \begin{align}
        u(x_i,t+1) = u(x_i, t) + dt \ \alpha \frac{u(t,x_{i-1}) - 2 u(t,x_i) + u(t,x_{i+1})}{2h}
    \end{align}
    We did not specify the how to generate the grid, \emph{i.e.}\ equidistant nodal spacing with $n$ grid points $x = \{ x_i = i \cdot h \in \mathbb{R} \vert i = 0,\ldots,n-1\} $, nor what time integration method to use, \emph{e.g.}\ the Euler method.
\end{enumerate}
Whether the language was C\texttt{++}, Fortran, Go, Julia, Matlab, Python, R, and Rust, we copied the generated code to the paper's GitHub repo\footnote{\url{https://github.com/diehlpkpapers/heat-ai/tree/main}}. For some of the generated codes, ChaptGPT provided instructions on how to compile the code. This instruction was added to the \textit{Readme.md}.

\section{Quality of the generated software}
\label{sec:quality}
In this section, we assess the quality of the AI-generated software. First, we checked whether the code compiles with a recent compiler.
Second, we checked whether the code executed without segmentation faults or other runtime errors. Third, we checked whether the code produced a correct result. The results for our test cases were the following:
\begin{itemize}
    \item[] $\int\limits_{-\pi}^{2/3\pi}  \sin(x) dx = -\cos(2/3\pi)+cos(-\pi) = -0.5$
    \item[] We solve $M \times x = b$ with the following values $A = \begin{pmatrix} 4 & 1 \\ 1 & 3 \end{pmatrix}$ and $b=\begin{pmatrix}
        1 \\ 2
    \end{pmatrix}$. The solution $x$ is $\begin{pmatrix} 0.09090909 & 0.63636364 \end{pmatrix}^T$.
\end{itemize}

\noindent Table~\ref{tab:code:quality} summarizes the results for the first two examples (NI and CGS). All codes and hand-repaired codes are available on GitHub\footnotemark[1]. However, for the sake of space we can not show them in the paper. We added the compilation and runtime errors to GitHub as well.
The results are shown in Table~\ref{tab:code:quality:parallel}. Table~\ref{tab:code:versions} lists the versions of the tools used.\\

\begin{table}[tb]
    \centering
    \caption{Three aspects of code quality: Compilation (Did the code compile with a recent compiler?), Runtime (Did the code execute without errors?), and Correctness (Did the code compute correct results?).}
    \label{tab:code:quality}
    \begin{tabular}{lc|cccccccccc}\toprule
    Attribute & V & C++ &  Python & Julia & Java & Go & Rust & Fortran & Matlab & R \\\midrule
    \multicolumn{11}{c}{Numerical integration (\textbf{NI})} \\\midrule
      Compila- &  3.5 & \tikzcmark   & -- & -- & \tikzcmark  &  \tikzcmark & \tikzcmark & \tikzxmark  &--  & -- \\
    tion & 4.0 & \tikzcmark  & -- & -- &  \tikzcmark & \tikzcmark & \tikzcmark & \tikzcmark & -- & -- \\\midrule
    Runtime   &  3.5 & \tikzcmark  & \tikzcmark  &  \tikzcmark  & \tikzcmark &  \tikzcmark & \tikzcmark & \tikzcmark & \tikzcmark & \tikzcmark  \\
    & 4.0 & \tikzcmark   & \tikzxmark  &  \tikzcmark &  \tikzcmark & \tikzcmark & \tikzcmark  & \tikzcmark &\tikzcmark  & \tikzcmark \\\midrule
     Correct-   &  3.5 &  \tikzcmark   & \tikzcmark &  \tikzcmark & \tikzcmark &  \tikzcmark  & \tikzcmark & \tikzcmark & \tikzcmark & \tikzcmark \\
    ness& 4.0 & \tikzcmark & \tikzxmark &  \tikzxmark & \tikzxmark  & \tikzxmark & \tikzxmark & \tikzxmark &\tikzxmark  & \tikzxmark \\\midrule
    \multicolumn{11}{c}{Conjugate gradient solver (\textbf{CGS})} \\\midrule
    Compila- &  3.5 & \tikzcmark  & -- & -- & \tikzcmark & \tikzcmark & \tikzxmark & \tikzxmark & --   & -- \\
    tion & 4.0 & \tikzcmark   & -- & -- & \tikzcmark & \tikzcmark & \tikzcmark& \tikzcmark  &  -- & -- \\\midrule
    Runtime   &  3.5 & \tikzcmark  & \tikzcmark & \tikzcmark   & \tikzcmark & \tikzcmark & \tikzcmark & \tikzcmark & \tikzcmark & \tikzcmark \\
    & 4.0 & \tikzcmark   & \tikzcmark & \tikzxmark  & \tikzcmark &  \tikzcmark & \tikzcmark & \tikzcmark & \tikzcmark  &\tikzxmark \\\midrule
     Correct-   &  3.5 & \tikzcmark   & \tikzcmark & \tikzcmark & \tikzcmark & \tikzxmark  & \tikzcmark & \tikzcmark & \tikzcmark  & \tikzcmark \\
    ness& 4.0 & \tikzcmark  & \tikzcmark & \tikzcmark &  \tikzcmark &  \tikzcmark & \tikzcmark & \tikzcmark &\tikzcmark  & \tikzxmark \\\bottomrule
    \end{tabular}
\end{table}

\begin{table}[tb]
    \centering
        \caption{Three aspects of code quality for the \textbf{parallel heat equation solver (PHS)}: Compilation (Did the code compile with a recent compiler?), Runtime (Did the code execute without errors?), and Correctness (Did the code compute correct results?).}
    \label{tab:code:quality:parallel}
      \begin{tabular}{lc|cccccccccccc}\toprule
  Attribute & V & C++ & Kokkos & HPX & Python & Julia & Java & Go & Rust & Fortran & Matlab & R \\\midrule
      Compila- &  3.5 & \tikzcmark & \tikzcmark & \tikzcmark & -- & -- & \tikzcmark & \tikzcmark & \tikzxmark & \tikzxmark & -- & -- \\
    tion & 4.0 & \tikzxmark & \tikzcmark & \tikzcmark & -- & -- & \tikzcmark & \tikzcmark & \tikzcmark & \tikzcmark & -- & -- \\\midrule
    Runtime   &  3.5 & \tikzcmark &  \tikzcmark& \tikzxmark & \tikzxmark & \tikzcmark & \tikzxmark & \tikzxmark & & \tikzcmark &\tikzcmark &\tikzxmark  \\
    & 4.0 & \tikzcmark & \tikzcmark & \tikzcmark & \tikzxmark & \tikzxmark & \tikzcmark & \tikzxmark & \tikzcmark & \tikzcmark &\tikzcmark  &\tikzcmark\\\midrule
     Correct-   &  3.5 & \tikzcmark &  \tikzxmark& \tikzcmark & \tikzxmark & \tikzxmark & \tikzcmark & \tikzxmark & & \tikzxmark &\tikzcmark &\tikzcmark \\
    ness& 4.0 & \tikzcmark & \tikzxmark & \tikzxmark & \tikzcmark & \tikzcmark&\tikzcmark & \tikzxmark & \tikzxmark & \tikzxmark &\tikzcmark  &\tikzcmark\\\bottomrule
\end{tabular}

\end{table}

\begin{table}[tb]
    \centering
    \caption{Version of all used tools in this study}
    \label{tab:code:versions}
    \begin{tabular}{l|cccccccccccc}\toprule
     Tool & C++ &  Python & Julia & Java & Go & Rust & Fortran & Matlab & R & Kokkos & HPX \\\midrule
     Version & 12 & 3.9.18 & 1.10.3 & 22 & 1.20.12 & 1.70 & 12 & R2024a & 4.4.0 & & 1.9.1 \\\bottomrule
\end{tabular}

\end{table}

\noindent\textbf{C\texttt{++}} The trapezoidal rule was used for the numerical integration. Both codes finished and returned the correct result. The ChaptGPT\ $4.0$ code computed two results once using the trapezoidal rule, with and without absolute values of the function. For the conjugate gradient solver both codes compiled, finished, and produced correct results. For, the parallel heat equation solver, OpenMP was used for parallelism. The code generated by ChatGPT\ $3.5$ compiled, the code generated by ChatGPT\ $4.0$ had issues with shared variables in the OpenMP parallel region (\bash{error: r is predetermined shared for shared}). Both codes had no runtime errors and produced plausible results.\\

\noindent\textbf{Kokkos} The ChatGPT\ $3.5$ version and ChatGPT\ $4.0$ version used a \cpp{Kokkos::parallel_for} loop for parallel computing. Both codes were compiled with the latest Kokkos version using \texttt{CMake}. Both codes produced incorrect results.\\

\noindent\textbf{HPX}
The generated code for ChatGPT\ $3.5$ included some HPX-specific headers but none of the parallel constructs. The code generated by ChatGPT\ $4.0$ used HPX's parallel algorithms for parallelism. Both codes compiled after small changes in header names and namespaces due to the HPX version. ChatGPT did not mention which HPX version to use. All values for the ChatGPT\ $4$ version produced \cpp{NaN}.\\

\noindent\textbf{Python}
In the NI example, the code generated by ChatGPT\ $3.5$ ran without runtime errors and produced accurate results. ChatGPT\ $4.0$ provided two codes, one utilizing the numpy library and the other employing the scipy library. Upon testing both, a runtime error occurred in the numpy-based code due to a \texttt{TypeError: bad operand type}. Unfortunately, neither code option from ChatGPT\ $4.0$ produced correct results.
In the CGS example, both code snippets ran without runtime errors and yielded correct results.
In the PHS example, the code generated by ChatGPT\ $3.5$ utilized the \textit{joblib} library to implement parallelism, whereas the code from ChatGPT\ $4.0$ used the multiprocessing library. Although both pieces of code were syntactically correct and valid Python code, they each encountered runtime errors. However, while the errors in the ChatGPT\ $3.5$-generated code remain unresolved, the issues in the ChatGPT\ $4.0$-generated code were promptly addressed, and provided correct results.\\

\noindent\textbf{Julia}
For the numerical integration, the ChatGPT\ $4.0$ version computed $3.5$ for the integral by using the absolute value of the function for the trapezoidal rule.
For the conjugate gradient solver, the ChatGPT\ $4.0$ code stopped with \texttt{ERROR: LoadError: UndefVarError: `dot` not defined}. After copying the dot product function from the ChatGPT\ $3.5$ version, the code worked. Both codes returned the correct result. For the parallel heat equation solver, the ChatGPT\ $3.5$ version produced final values were zero (which was incorrect). A short time later, the code stopped with the error \texttt{Unhandled Task ERROR: On worker 2: UndefVarError: `u` not defined}. The ChatGPT $4.0$ code did not print anything and crashed with \texttt{KeyError: key SharedArrays [1a1011a3-84de- 559e-8e89-a11a2f7dc383] not found}. The ChatGPT\ $3.5$ version used a for loop for parallelism (\lstinline[language=python,breaklines=true]{@distributed}) and ChaptGPT\ 4 used \texttt{@spawnat} for parallelism. The ChatGPT\ $3.5$ version incorrectly parallelized the loop for the time steps instead of the loop for the domain. The ChatGPT\ $4.0$ version added code to plot the results, however, we changed the code to print the results.\\

\noindent\textbf{Java} 
For the NI example, both codes successfully compiled and ran. The code generated by ChatGPT\ $3.5$ produced accurate results, while the code generated by ChatGPT\ $4.0$ did not. In the CGS example, both codes successfully compiled, ran, and produced correct results. 
In the PHS example, both the generated codes used Java's built-in concurrency utilities, particularly the \lstinline[language=java]{ExecutorService} from the \lstinline[language=java]{java.util.concurrent} package. This utility manages a pool of threads and allows user to execute tasks in parallel. Both codes compiled, but the ChatGPT\ $3.5$-generated code encountered runtime errors. This error was \texttt{Index Out Of Bounds Exception} which we  fixed. The code generated by ChatGPT\ $4.0$ had no errors.\\

\noindent\textbf{Go}
For the numerical integration, ChatGPT\ $4.0$ computed the integral as $3.5$ while using absolute values for the function evaluation. Everything else was fine. For the conjugate gradient solver all codes worked. For the parallel heat equation, the ChatGPT\ $3.5$ version of the code compiled after fixing the error \bash{"math" imported and not used} by removing the unused package. The ChatGPT\ $4.0$ compiled without issue. The ChatGPT\ $3.5$ code crashed with the error \bash{panic: runtime error: index out of range [-1] with length 101}. After fixing the the code by hand, it crashed with the error \bash{panic: runtime error: index out of range [101] with length 101}. The final values were zero or one and the result was not correct. The ChatGPT\ $4.0$ code crashed with the error \bash{panic: sync: negative WaitGroup counter}. After fixing the error most of the final values were not-a-number (\cpp{NaN}).\\

\noindent\textbf{Rust}
For the numerical integration, the ChatGPT\ $4.0$ version evaluated the integral as $3.5$. Everything else worked for both codes. For the conjugate gradient solver, the ChatGPT\ $3.5$ version did not compile because \rust{extern crate special;} was missing at the top of the \bash{main.rs}. After fixing that problem, both codes executed and provided the correct result. For the parallel heat equation solver, both codes used the data-parallelism library Rayon for parallelism. 
The code for the ChatGPT\ $4.0$ version compiled, however, \rust{extern crate rayon;} was missing. However, some of the values were negative which is not correct.\\

\noindent\textbf{Fortran}
For the numerical integration, the ChatGPT\ $3.5$ code did not compile due to \cpp{Error: EXTERNAL attribute conflicts with FUNCTION attribute}. The ChatGPT\ $4.0$ code computed the integral as $3.5$ by using absolute values for the function evaluation for the trapezoidal rule. For the conjugate gradient solver, the ChatGPT\ $3.5$ version did not compile since there was a variable \lstinline[language=fortran]{dot_product} and a function \lstinline[language=fortran]{dot_product} defined. After fixing this error by hand, both codes compiled and provided correct results. For the parallel heat equation solver, OpenMP was used by ChaptGPT\ $3.5$, and Fortran coarray by ChatGPT\ $4.0$. Neither of these codes compiled. The errors were \bash{Error: Symbol t at (1) already has basic type of REAL} and \bash{Error: !\$OMP PARALLEL DO iteration variable must be of type integer at (1)}. After fixing these compilation errors by hand, we had to install Fortran coarray. Neither code produced correct results.\\

\noindent\textbf{Matlab} In the NI example, both codes ran without runtime errors. The code generated by ChatGPT\ $3.5$ yielded accurate results, whereas the code generated by ChatGPT\ $4.0$ failed to produce correct results. For the conjugate gradient solver, both codes ran without errors and produced accurate results. In the PHS example, the code generated by ChatGPT\ $3.5$ run without errors, but it did not utilize parallel computing. In contrast, the code generated by ChatGPT\ $4.0$ effectively used the Parallel Computing Toolbox, ran without errors, and produced correct results.\\

\noindent\textbf{R} 
As presented in table \ref{tab:code:quality}, both codes ran without runtime errors for the numerical integration. The code generated by ChatGPT\ $3.5$ produced accurate results, while the code generated by ChatGPT\ $4.0$ did not produce correct results. In the CGS example, the ChatGPT\ $3.5$-code ran without errors and produced correct results. However, runtime errors occurs with the ChatGPT\ $4.0$-generated code, and these remain unaddressed.
In the PHS example, the ChatGPT\ $3.5$-generated code used loops without parallelism since the loops were relatively small and the computations were not too expensive. The originally-generated code encountered an error related to a missing package ('\textit{animation}'). To fix this error, we used a different approach and plotted the output, the results were accurate. The code generated with ChatGPT 4 used the \textit{parallel} library and produced correct results.

\subsection{Common issues}
Here we present the list of common issues which we observed while debugging:
\begin{itemize}
    \item \textbf{Compilation}: \\
    For NI and CGS, most compilation errors were minor and could be easily fixed. For PHS, most of the errors were related to parallelism, and sometimes knowledge about the parallel programming language or library was needed to fix the problem. In total, we had five compilation errors for all examples.
    \item \textbf{Runtime}: \\
    Most of the runtime errors were minor and could be easily fixed. Most were type errors, undefined variables, and undefined functions for interpreted languages. Other errors were index out-of-bound exceptions for the heat equation solver for the first and last element with the stencil. Some errors were related to the parallelism and knowledge about the parallel programming language or library was required to address them.
    \item \textbf{Correctness}: \\
    For NI, the ChatGPT\ $3.5$ versions produced all the correct results. The ChatGPT\ $4.0$ version computed the integral as $3.5$ because it used absolute values for the function evaluation. After removing the absolute values all codes computed the correct result. For the conjugate gradient solver, all except two codes produced the correct result. For the single-threaded codes, overall most results were correct. For the parallel codes, $11$ codes produced correct results, and $10$ codes did not.
\end{itemize}

\section{Code metrics}
\label{sec:code_metrics}
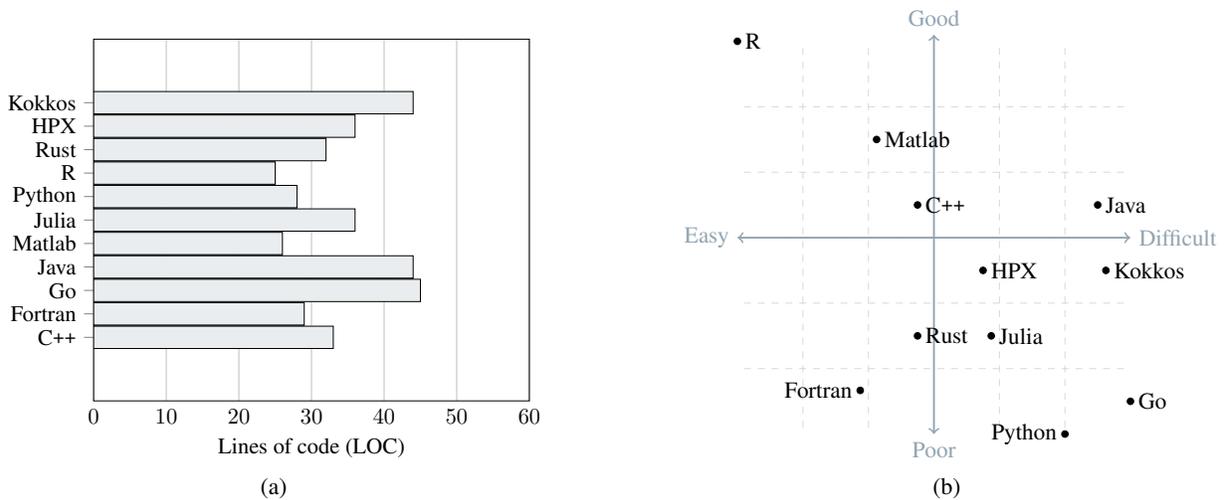
\begin{figure}[tb]
    \centering
       \subfloat[\label{fig:line:of:codes}]{

 \resizebox{0.45\textwidth}{!}
        {
  \begin{tikzpicture}
 \begin{axis}[
    xbar=12pt,
    xmin=0,xmax=60,
    ytick=data,
    enlarge y limits={abs=1cm},
    symbolic y coords={C\texttt{++},Fortran,Go,Java,Matlab,Julia,Python,R,Rust,HPX,Kokkos},
    bar width = 10pt,
    xlabel= Lines of code (LOC), 
    ytick align=outside, 
    ytick pos=left,
    major x tick style ={ transparent},
    legend style={at={(0.04,0.96)},anchor=north west, font=\footnotesize, legend cell align=left},
    xmajorgrids=true
        ]    
    \addplot[xbar,fill=cadetgrey!20, area legend] coordinates {
        (33,C\texttt{++})
        (29,Fortran)
        (45,Go)
        (44,Java)
        (26,Matlab)
        (36,Julia)
        (28,Python)
        (25,R)
        (32,Rust)
        (36,HPX)
        (44,Kokkos)
        };
\end{axis}
\end{tikzpicture}
        }
    }
\hfill
    \subfloat[\label{fig:two:dim:plot} ]{
 \resizebox{0.45\textwidth}{!}
        {
 \begin{tikzpicture}
    \draw[help lines, color=gray!30, dashed] (-2.9,-2.9) grid (2.9,2.9);
    \draw[<->,thick,cadetgrey] (-3,0)--(3,0) node[right]{Difficult};
    \draw[<->,thick,cadetgrey] (0,-3)--(0,3.1) node[above,cadetgrey]{Good};
    \node[left,cadetgrey] at (-3,0) {Easy};
    \node[below,cadetgrey] at (0,-3) {Poor};
    \draw[fill=black] (2,-3) circle [radius=0.05] node[left] {Python};
    \draw[fill=black] (2.625,-0.5) circle [radius=0.05] node[right] {Kokkos};
    \draw[fill=black] (-0.25,0.5) circle [radius=0.05] node[right] {C\texttt{++}};
    \draw[fill=black] (-1.125,-2.333333333) circle [radius=0.05] node[left] {Fortran};
    \draw[fill=black] (3,-2.5) circle [radius=0.05] node[right] {Go};
    \draw[fill=black] (0.75,-0.5) circle [radius=0.05] node[right] {HPX}; 
    \draw[fill=black] (2.5,0.5) circle [radius=0.05] node[right] {Java};   
    \draw[fill=black] (0.875,-1.5) circle [radius=0.05] node[right] {Julia};  
    \draw[fill=black] (-0.875,1.5) circle [radius=0.05] node[right] {Matlab};  
    \draw[fill=black] (-3,3) circle [radius=0.05] node[right] {R};  
    \draw[fill=black] (-0.25,-1.5) circle [radius=0.05] node[right] {Rust};  
    \end{tikzpicture}
        }}
    \caption{Software engineering metrics for the parallel heat equation solver: \protect\subref{fig:line:of:codes} Lines of code for all implementations using the maximal lines of code. The numbers were determined with the Linux tool \textit{cloc} and   \protect\subref{fig:two:dim:plot} Two-dimensional classification using the computational time and the COCOMO model.}
\end{figure}

The lines of code were determined with the Linux tool \texttt{cloc} and the larger amount of code lines between the ChaptGPT\ $3.5$ and ChatGPT\ $4.0$ were chosen. The difference was between one and five lines of code. The lines of code were in the same ballpark for all of the codes.
The lines of code metric, however, does not measure quality well. For this, we use the \textbf{Co}nstructive \textbf{Co}st \textbf{Mo}del (COCOMO)~\cite{stutzke1997software,barry1981software}. COCOMO is a general model with no specialization for parallel programming.
There were some attempts to add this capability for the COCOMO\ II model~\cite{miller2018applicability}. Until we get a specialized model for parallel computing or HPC, the COCOMO model is a reasonable candidate for measuring quality.
We use the tool \textit{Sloc Cloc and Code} (scc)\footnote{\url{https://github.com/boyter/scc}} to get the COCOMO metrics for each approach. We use the average of the COCOMO metric for version $3.5$ and version $4.0$ for the classification of \textbf{easy} and \textbf{difficult}. We use the three quantities of interest in Table~\ref{tab:code:quality} for the code quality. We use the following metric
\begin{align}
    q(\text{language}) := \frac{1}{2}\left(\frac{\text{comp}+\text{run}+\text{correct}}{3} + \frac{\text{comp}+\text{run}+\text{correct}}{3}\right)
\end{align}
to classify the code quality from \textbf{poor} to \textbf{good}. Figure~\ref{fig:line:of:codes:int} shows the lines of codes for the numerical integration. Here, Matlab and R have the least lines of code. Python and Julia are comparable. Java, Go, Fortran, and C\texttt{++} have comparable results. Figure~\ref{fig:two:dim:plot:int} shows the two-dimensional classification. Concerning the code quality, Matlab, Python, Java, and C\texttt{++} showed the best results. Fortran, Go, and Rust, are very close for code quality and complexity. Julia and R are last. 
\begin{figure}[tb]
    \centering
       \subfloat[\label{fig:line:of:codes:int}]{

 \resizebox{0.45\textwidth}{!}
        {
  \begin{tikzpicture}
 \begin{axis}[
    xbar=12pt,
    xmin=0,xmax=40,
    ytick=data,
    enlarge y limits={abs=1cm},
    symbolic y coords={C\texttt{++},Fortran,Go,Java,Matlab,Julia,Python,R,Rust},
    bar width = 10pt,
    xlabel= Lines of code (LOC), 
    ytick align=outside, 
    ytick pos=left,
    major x tick style ={ transparent},
    legend style={at={(0.04,0.96)},anchor=north west, font=\footnotesize, legend cell align=left},
    xmajorgrids=true
        ]    
    \addplot[xbar,fill=cadetgrey!20, area legend] coordinates {
        (29,C\texttt{++})
        (23,Fortran)
        (25,Go)
        (21,Java)
        (5,Matlab)
        (12,Julia)
        (10,Python)
        (6,R)
        (23,Rust)
        };
\end{axis}
\end{tikzpicture}
        }
    }
    \subfloat[\label{fig:two:dim:plot:int} ]{
 \resizebox{0.45\textwidth}{!}
        {
 \begin{tikzpicture}
    \draw[help lines, color=gray!30, dashed] (-2.9,-2.9) grid (2.9,2.9);
    \draw[<->,thick,cadetgrey] (-3,0)--(3,0) node[right]{Difficult};
    \draw[<->,thick,cadetgrey] (0,-3)--(0,3.1) node[above,cadetgrey]{Good};
    \node[left,cadetgrey] at (-3,0) {Easy};
    \node[below,cadetgrey] at (0,-3) {Poor};
    \draw[fill=black] (-1.094117647,3) circle [radius=0.05] node[right] {Python};
    \draw[fill=black] (3,3) circle [radius=0.05] node[right] {C\texttt{++}};
    \draw[fill=black] (1.870588235,1) circle [radius=0.05] node[left] {Fortran};
    \draw[fill=black] (2.294117647,1) circle [radius=0.05] node[right] {Go};
    \draw[fill=black] (1.517647059,3) circle [radius=0.05] node[right] {Java};   
    \draw[fill=black] (-0.6,0) circle [radius=0.05] node[right] {Julia};  
    \draw[fill=black] (-3,3) circle [radius=0.05] node[right] {Matlab};  
    \draw[fill=black] (-2.576470588,-3) circle [radius=0.05] node[above] {R};  
    \draw[fill=black] (2.011764706,1) circle [radius=0.05] node[below] {Rust};  
    \end{tikzpicture}
        }}
    \caption{Software engineering metrics for the \textbf{numerical integration example}: \protect\subref{fig:line:of:codes:int} Lines of code for all implementations using the maximal lines of code. The numbers were determined with the Linux tool \textit{cloc} and   \protect\subref{fig:two:dim:plot:int} Two-dimensional classification using the computational time and the COCOMO model.}
\end{figure}
Figure~\ref{fig:line:of:codes:cgm} shows the lines of code for the conjugate gradient method. Here, R, Fortran, Julia, and Matlab have comparable results. Fortran and Rust are close. Java, Go, Fortran, and C\texttt{++} are comparable. Figure~\ref{fig:two:dim:plot:cgm} shows the two-dimensional classification. Here, Go, Java, and C\texttt{++} are on the upper end for the code quality. Fortran and Rust are in the middle for code quality and on the opposite side for complexity. Julia and Matlab are second last and Python and R are last.
\begin{figure}[tb]
    \centering
           \subfloat[\label{fig:line:of:codes:cgm}]{

 \resizebox{0.45\textwidth}{!}
        {
  \begin{tikzpicture}
 \begin{axis}[
    xbar=12pt,
    xmin=0,xmax=80,
    ytick=data,
    enlarge y limits={abs=1cm},
    symbolic y coords={C\texttt{++},Fortran,Go,Java,Matlab,Julia,Python,R,Rust},
    bar width = 10pt,
    xlabel= Lines of code (LOC), 
    ytick align=outside, 
    ytick pos=left,
    major x tick style ={ transparent},
    legend style={at={(0.04,0.96)},anchor=north west, font=\footnotesize, legend cell align=left},
    xmajorgrids=true
        ]    
    \addplot[xbar,fill=cadetgrey!20, area legend] coordinates {
        (67,C\texttt{++})
        (46,Fortran)
        (72,Go)
        (69,Java)
        (36,Matlab)
        (37,Julia)
        (30,Python)
        (31,R)
        (46,Rust)
        };
\end{axis}
\end{tikzpicture}
        }
    }
        \subfloat[\label{fig:two:dim:plot:cgm} ]{
 \resizebox{0.45\textwidth}{!}
        {
 \begin{tikzpicture}
    \draw[help lines, color=gray!30, dashed] (-2.9,-2.9) grid (2.9,2.9);
    \draw[<->,thick,cadetgrey] (-3,0)--(3,0) node[right]{Difficult};
    \draw[<->,thick,cadetgrey] (0,-3)--(0,3.1) node[above,cadetgrey]{Good};
    \node[left,cadetgrey] at (-3,0) {Easy};
    \node[below,cadetgrey] at (0,-3) {Poor};
    \draw[fill=black] (-1.909090909,-3) circle [radius=0.05] node[left] {Python};
    \draw[fill=black] (1.363636364,2.820895522) circle [radius=0.05] node[right] {C\texttt{++}};
    \draw[fill=black] (-3,-0.044776119) circle [radius=0.05] node[right] {Fortran};
    \draw[fill=black] (-1.909090909,2.910447761) circle [radius=0.05] node[right] {Go};
    \draw[fill=black] (0.272727273,3) circle [radius=0.05] node[right] {Java};   
    \draw[fill=black] (-0.272727273,-1.656716418) circle [radius=0.05] node[right] {Julia};  
    \draw[fill=black] (1,-1.298507463) circle [radius=0.05] node[right] {Matlab};  
    \draw[fill=black] (-0.272727273	,-2.731343284) circle [radius=0.05] node[above] {R};  
    \draw[fill=black] (3,-0.134328358) circle [radius=0.05] node[left] {Rust};  
    \end{tikzpicture}
        }}
    \caption{Software engineering metrics for the \textbf{conjugate gradient solver}: \protect\subref{fig:line:of:codes:cgm} Lines of code for all implementations using the maximal lines of code. The numbers were determined with the Linux tool \textit{cloc} and   \protect\subref{fig:two:dim:plot:cgm} Two-dimensional classification using the computational time and the COCOMO model.}
\end{figure}
Figure~\ref{fig:line:of:codes} shows the lines for the parallel heat equation solver. Here, Kokkos, Java, and Go are at the top. HPX, Python, Rust, C\texttt{++}, and Julia are in the middle. R, Matlab, and Fortran are at the lower end. Figure~\ref{fig:two:dim:plot} shows the classification of the parallel heat equation solver. Matlab resulted in the best code quality. Next, R, C\texttt{++}, and Java had a similar code quality. Looking at GitHub C\texttt{++}, Python, and Java were in the top ten most used languages in 2022. Maybe the larger training data set explains why C\texttt{++} and Java had good results. However, Python is an outlier. It is the second most used language on GitHub, but the code quality was low.

\section{Discussion and Conclusion}
\label{sec:conclusion}
In this work we have conducted an evaluation of three computational problems using ChatGPT versions $3.5$ and $4.0$ for code generation using a range of programming languages. We evaluated the compilation, runtime errors, and accuracy of the codes that were produced. We tested their accuracy, first with a basic numerical integration, the with a conjugate gradient solver, and finally with a 1D stencil-based heat equation solver. 

For the numerical integration example, codes generated by both versions compiled successfully in all languages except Fortran, and executed without any runtime errors. However, the accuracy of the outputs from the ChatGPT\ $4.0$-generated codes was incorrect, possibly due to the misinterpretation of the keyword ``area'' in the prompt. In the case of the conjugate gradient solver, all generated codes compiled successfully with the exceptions of Fortran and Rust. Despite these compilation issues, the resultant codes from all other languages produced correct results, except for R.
The parallel 1D heat problem proved to be the most challenging for the AI. Compilation errors were noted in the codes for Fortran, Rust, and C\texttt{++}. Furthermore, a majority of the generated codes encountered runtime errors, and most failed to produce correct results, indicating substantial issues with the implementation logic or the handling of parallel computing constructs by the AI code generator models.

We then analyzed the lines of code for all the generated codes, and the code quality using the COCOMO metric. The analysis of lines of code across all examples showed that Matlab and R consistently produced the lower lines of codes values, followed by Python, Julia, and Fortran (Section~\ref{sec:code_metrics}). In terms of code quality, C\texttt{++} and Java consistently demonstrated robustness across all the examples tested, followed by Matlab. These languages appear to offer a balance between code quality and complexity, making them suitable choices for more complex computational tasks. 

For future work, we plan to study whether ChaptGPT can be made to fix its code bugs by submitting new requests. Another follow-up would be to study how well ChatGPT can produce code for native GPU kernels. Note that the Kokkos codes we have generated can already run on the GPU. Another direction for study would be to evaluate the performance of the AI-generated code. Lastly, prompting it to create distributed code versions would be interesting.       

\subsection*{Acknowledgments}
{This work was supported by the U.S. Department of Energy through the Los Alamos National Laboratory. Los Alamos National Laboratory is operated by Triad National Security, LLC, for the National Nuclear Security Administration of U.S. Department of Energy (Contract No. 89233218CNA000001).}

\printbibliography

\end{document}